\documentclass[%
reprint,
superscriptaddress,
amsmath,amssymb,
aps,
prb,
floatfix,
]{revtex4-2}
\usepackage[normalem]{ulem}
\usepackage[english]{babel}
\usepackage{graphicx}
\usepackage{graphicx}
\usepackage{dcolumn}
\usepackage{booktabs}
\usepackage{bm}
\usepackage{braket}
\usepackage[colorlinks]{hyperref}
\usepackage[usenames,dvipsnames]{xcolor}
\usepackage{siunitx}
\usepackage{breqn}
\definecolor{darkblue}{RGB}{0,32,96}
\usepackage[dvipsnames]{xcolor}
\usepackage[version=3]{mhchem}

\emergencystretch=\maxdimen
\newcommand{\be}{\begin{equation}}
\newcommand{\ee}{\end{equation}}
\renewcommand{\phi}{\varphi}

\begin{document}

\title{Negative non-linear differential dissipation in mechanical quartz resonators }

\newcommand{\Aalto}{\affiliation{\mbox{Low Temperature Laboratory, Department of Applied Physics, Aalto University, P.O. Box 15100, FI-00076 AALTO, Finland}}}
\newcommand{\QTFce}{\affiliation{\mbox{QTF Centre of Excellence, Department of Applied Physics, Aalto University, P.O. Box 15100, FI-00076 AALTO, Finland}}}
\newcommand{\USEN}{\affiliation{\mbox{Department of Science and Industry Systems, University of South-Eastern Norway, P.O. Box 235, Kongsberg, Norway}}}
\newcommand{\UiO}{\affiliation{\mbox{Department of Physics, University of Oslo, Oslo N-0316, Norway}}}
\newcommand{\UPOL}{\affiliation{Department of Optics, Palacký University, 17. listopadu 1192/12, 77900 Olomouc, Czechia}}

\author{Elena Sergeicheva}
\Aalto \QTFce

\author{Juuso Manninen}
\USEN \UPOL

\author{Juha Tuoriniemi}
\Aalto \QTFce

\author{Weijun Zeng}
\Aalto \QTFce

\author{Francesco Massel}
\USEN \UiO

\author{Pertti Hakonen}
\Aalto \QTFce

\date{\today}

\begin{abstract}
We observe negative non-linear differential dissipation in a low-frequency (32.7 kHz) miniature quartz tuning fork, 
representing a high-Q piezoelectric resonator, at temperatures below 0.3 K.
Similar drive-induced dissipation suppression has previously been reported in superconducting microwave resonators, 
and high-frequency MEMS devices, associated to coupling with tunneling two-level systems (TTLSs).
In the case of low-frequency resonators, the equivalent temperature of the mechanical mode is orders of magnitude lower 
than the thermodynamic temperature of the device, implying that accessible TTLS states are largely saturated.
We demonstrate that the observed temperature and drive dependence of the losses fit well within the framework of an interacting TTLS bath, 
indicating the presence of a channel for energy transfer between the mechanical mode and thermally activated TTLSs.
As a possible coupling mechanism, we suggest microwave phonon modes in the metallic electrodes of the tuning fork.
\end{abstract}
\maketitle

\section{Introduction}
\label{sec:intro}

Low-loss mechanical resonators are central to precision force sensing~\cite{rugar_single_2004,moser_ultrasensitive_2013}, frequency metrology~\cite{grop_unprecedented_2011,ivanov_low_2009}, and cryogenic measurements~\cite{cleland_nanometre-scale_1998,lahaye_approaching_2004}, and are increasingly being used as interfaces on hybrid quantum platforms~\cite{bochmann_nanomechanical_2013,andrews_bidirectional_2014}. In many of these applications, performance is ultimately set by dissipation and its dependence on temperature and drive, making it essential to identify the dominant loss channels and their microscopic origin~\cite{ekinci_nanoelectromechanical_2005}.

The dissipation mechanism in mechanical resonators can be ascribed to multiple factors, among which clamping losses often provide a major contribution~\cite{safavi-naeini_two-dimensional_2014,tsaturyan_ultracoherent_2017}. A range of engineering approaches has therefore been developed to suppress energy leakage into the supports, including phononic-bandgap shielding and related acoustic isolation concepts~\cite{safavi-naeini_two-dimensional_2014,tsaturyan_ultracoherent_2017}.
In high-stress devices, dissipation dilution can further enhance the quality factor by storing most of the mechanical energy in a nearly lossless tensile field rather than in bending~\cite{tsaturyan_ultracoherent_2017}.

Even when such extrinsic channels are strongly reduced, intrinsic material losses can still dominate. In particular, the presence of two-level systems (TLSs) strongly influences dissipation in high-quality mechanical resonators. TLSs  are a phenomenological way to model degrees of freedom that can switch between two nearly degenerate arrangements and couple to strain and electric fields. They were introduced~\cite{anderson_anomalous_1972,Phillips.1987} to explain universal low‑temperature anomalies (specific heat, thermal conductivity, acoustic loss) in glassy materials and have been invoked to model losses in various setups, ranging from mechanical dissipation in MEMS/NEMS oscillators ~\cite{mohanty_intrinsic_2002,atalaya_nonlinear_2016} to dielectric losses and 1/{\itshape f} noise in superconducting qubits and resonators~\cite{martinis_decoherence_2005}.

In our work, we  investigate the mechanical properties of a quartz tuning fork, with particular emphasis on its dissipation properties. Quartz is commonly used to make tuning forks operating in the kilohertz regime, most often at 32 kHz. These forks can be produced in geometries for which stability against thermal fluctuations is excellent. Although widely used for studies of bulk quantum fluids~\cite{Blaauwgeers2007,Bradley2010,Pentti2011,Bradley2012,Skyba2014,Riekki2019,Skyba2020,Skyba2020b} and their liquid vapor surface~\cite{Todo2017}, the low-temperature properties of forks in vacuum have not been discussed to the same extent.
To address this open question, we characterize a commercial 32 kHz quartz tuning fork operated in vacuum from 4 mK up to 1 K, combining continuous-wave spectroscopy with time-domain ring-down measurements. By varying the excitation amplitude over a wide range, we access both the linear response and the strongly driven nonlinear regime. We provide here a direct probe of amplitude- and temperature-dependent loss mechanisms in a low-frequency, high-Q piezoelectric resonator. In particular, the ring-down data allow us to verify that the drive-induced suppression of losses persists after the drive is switched off. This rules out explanations that rely on a continuously applied excitation (for example, steady-state heating and the associated thermoelastic backaction), and instead points to an internal energy-exchange mechanism that remains operative during free decay.

Our findings reveal a nontrivial temperature dependence of mechanical losses as a function of an external driving field. We attribute this behavior to the presence of TLSs. Following the standard description of TLS-induced losses~\cite{enss_low-temperature_2005,Behunin2016}, we model the dissipation as the combined contribution of a relaxation channel and a resonant-like channel that saturates at large oscillation amplitudes.

Unlike the conventional description of the resonant-like mechanism, in which the mechanical motion is resonantly coupled to TLSs, in our system the mechanical motion parametrically drives a resonant coupling between high-frequency two-level systems and an out-of-plane phononic mode of the metallic lead, whose frequency lies in the microwave regime.

\begin{figure}
    \centering
    \includegraphics[width=0.9\linewidth]{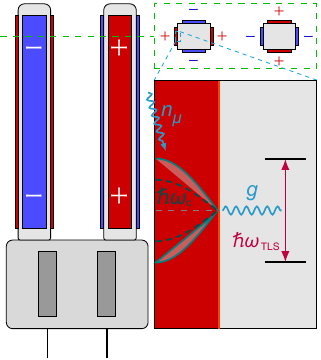}
    \caption{Depiction of the resonant-like coupling mechanism. The quartz tuning fork (gray) hosts an ensemble of TLSs (frequency $\omega_{\mathrm{TLS}}$) which are coupled resonantly to out-of-plane phonon mode (cavity mode) located across the Pd-based metallic lead on the fork (frequency $\omega_{\mathrm{c}}$). The mechanical displacement $x$, modulating $\omega_{\mathrm{TLS}}=\omega_{\mathrm{TLS}}(x)$, changes the detuning $\Delta(x) = \omega_{\mathrm{TLS}}(x) - \omega_{\mathrm{c}}$ between TLS and microwave mode. $n_\mu$ models the effective occupation of the microwave mode, induced by thermal radiation from the preamplifier far beyond its operating frequency.
    \label{fig:forkschematic}}
\end{figure}

A schematic setting of our parametric system is illustrated in Fig. \ref{fig:forkschematic}. 
The mechanical mode couples to the microwave phonon (MP) - resonant TLS ensemble by modulating the microwave detuning \footnote{Resonant microwave photons on the fork are excluded owing to the small size of the fork that yields resonances first above 10 GHz}. The resulting backaction on the mechanics is governed by the low-frequency susceptibility of the driven 
MP - TLS subsystem. 

\subsection{Resonant TLS channel as parametric backaction.}

Mechanical motion of the fork couples \emph{parametrically} to the TLS--MP-cavity subsystem by modulating the MP-cavity--TLS detuning $\Delta \doteq \omega_{\mathrm{TLS}} - \omega_{\mathrm{c}}$.
Such a modulation arises because the mechanical strain can shift the TLS level splitting (deformation-potential coupling), $\omega_{\mathrm{TLS}} \to \omega_{\mathrm{TLS}} + \delta \omega_{\mathrm{TLS}}(t)$. 
This leads to a time modulation of the detuning $ \delta\Delta(t) \doteq Gx(t)$, where the effective coupling constant $G$ characterizes the strength of the displacement-induced TLS frequency shift.

A minimal system Hamiltonian capturing this mechanism is
\begin{equation}
  \label{eq:1}
H =  \omega_{\mathrm{m}} b^\dagger b + \frac{\Delta(t)}{2}\sigma_z
+   g\left(a\sigma_+ + a^\dagger\sigma_-\right),
\end{equation}
where $b^\dagger (b)$ denotes the creation (annihilation) operator of fork quanta, $\Delta(t)=\Delta_{\mathrm{eq}} + Gx_{\rm zpf}(b^\dagger+b)$, and we have moved to a frame rotating at microwave phonon cavity resonant frequency $\omega_{\mathrm{c}}$ for both the cavity mode $a$ and for the TLS.

In our analysis, we assume that the cavity mode is populated through its coupling to a thermal environment. Correspondingly, dissipation and thermal occupation enter through the cavity damping rate and effective bath temperature. In this setting, the TLS--MP-cavity subsystem constitutes a dissipative, frequency-dependent environment for the mechanics. Since mechanical motion modulates detuning, it perturbs the steady state TLS population, which relaxes back to equilibrium with an effective response time $\tau_{\rm res}$.

From Eq.~\eqref{eq:1} we can write the average backaction force acting on the mechanics in a semiclassical picture as
\begin{equation}
  \label{eq:2}
  F=-\partial H/\partial x=-(1/2)G{\braket{\sigma_{z}}} .
\end{equation}
Using the Fourier convention $x(t)=\int (d\omega/2\pi)\,x(\omega)e^{-i\omega t}$, we obtain
\begin{equation}
  \label{eq:3}
  \delta F(\omega) = -\frac{G^2}{2}\chi_{zz}(\omega)x(\omega)
\end{equation}
Here $\chi_{zz}(\omega)$ is defined operationally via $ \delta {\braket{\sigma_{z}}} = \chi_{zz}(\omega) \delta \Delta$. In the low‑frequency limit $\omega \simeq \omega_{\mathrm m} \ll \omega_{\mathrm c}$, we model this 
response function phenomenologically by a single‑pole (Debye) form with an effective static amplitude $\chi_0(T)$ and relaxation time $\tau_{\mathrm{res}}$ as
\begin{equation}
  \label{eq:4}
  \chi_{zz}(\omega)\simeq \frac{\chi_0(T)}{1-i\omega\tau_{{\mathrm{res}}}}.
\end{equation}
We assume that $\tau_{\mathrm{res}}$ is approximately independent of temperature and oscillation amplitude over the measured range. At fixed mechanical frequency, the corresponding Debye factor can therefore be absorbed into the dissipation prefactor introduced below.
To leading order in the backaction, Eq.~\eqref{eq:3} gives the mechanical amplitude-decay rate
\begin{equation}
  \label{eq:5} \Gamma_{\mathrm{m,res}}(\omega)=-\frac{G^{2}}{4m\omega}\operatorname{Im}[\chi_{zz}(\omega)]
\end{equation}
and, correspondingly the frequency shift
\begin{equation}
  \label{eq:6} \delta\omega_{\mathrm{m,res}}(\omega)=\frac{G^{2}}{4m\omega}\operatorname{Re}[\chi_{zz}(\omega)].
\end{equation}
Given the expression for $ \chi_{zz}(\omega)$ shown in Eq.~\eqref{eq:4},
we can write Eqs.~\eqref{eq:5} and \eqref{eq:6} as
\begin{equation}
  \label{eq:7}
   \Gamma_{{\mathrm{m,res}}}(\omega) = -\frac{ G^{2}\chi_0(T)}{4m\omega}\,
  \frac{\omega\tau_{{\mathrm{res}}}}{1+\omega^{2}\tau_{{\mathrm{res}}}^{2}},
\end{equation}
\vspace{-8pt}
\begin{equation}
    \delta\omega_{\mathrm{m,res}}(\omega)=\frac{ G^{2}\chi_0(T)}{4m\omega}\,
\frac{1}{1+\omega^{2}\tau_{{\mathrm{res}}}^{2}}.
\end{equation}
Here $\chi_0(T)$ is the quasi-static ($\omega\to 0$) spin susceptibility of the TLS--MP-cavity system, and $\tau_{{\mathrm{res}}}$ is its effective relaxation (response) time.

\paragraph{Temperature dependence of the resonant channel.}
The resonant TLS contribution is set by the sub-ensemble of TLSs whose level splitting matches the localized cavity mode, $E\simeq E_{{\mathrm{res}}}$ (equivalently, small cavity--TLS detuning $\Delta \doteq \omega_{\mathrm{TLS}} - \omega_{\mathrm{c}}$). In this channel the longitudinal polarization $\langle\sigma_z\rangle=p_e-p_g$ changes because microwave-induced transitions transfer population between the TLS ground and excited states. For thermal TLS populations at temperature $T$,
\begin{align*}
    p_g(E,T) &= \frac{e^{+\beta E/2}}{e^{+\beta E/2}+e^{-\beta E/2}},\\
    p_e(E,T) &= \frac{e^{-\beta E/2}}{e^{+\beta E/2}+e^{-\beta E/2}},
\end{align*}
so that $\langle\sigma_z\rangle=-\tanh(\beta E/2)$, while net absorption is weighted by $p_g-p_e=-\langle\sigma_z\rangle$. A narrowband microwave phonon field with mean occupation $n_\mu$ induces upward (absorption) and downward (emission) transition rates which, in a Fermi--Golden--Rule description, can be written as
\[
\Gamma_{\uparrow}(\Delta)\propto p_g\,n_{\mu}\,L(\Delta),\qquad
\Gamma_{\downarrow}(\Delta)\propto p_e\,(n_{\mu}+1)\,L(\Delta),
\]
where $L(\Delta)$ is the resonant lineshape. For $n_\mu \gg 1$ (effective occupation excited by excess microwave back action from the cooled preamplifier), a small mechanical modulation $\delta\Delta(t)$ therefore modulates the transition rates through $L(\Delta)$,
so that the modulation of the net transition imbalance is weighted by the population difference,
\[
\delta \left[\Gamma_{\uparrow}-\Gamma_{\downarrow}\right]\propto (p_g-p_e) n_\mu (\partial_\Delta L) \delta\Delta(t)
\propto \tanh \left(\frac{\beta E}{2}\right)\delta\Delta(t).
\]
Together with the single-pole population response introduced above, this rate modulation motivates the same population-difference factor in $\chi_0(T)$, with the response time and the remaining coupling and detuning factors absorbed into $\tilde{\chi}_0$.
Restricting to the resonant TLS subset $E\simeq E_{{\mathrm{res}}}$ and absorbing the remaining (detuning-slope, geometry, and coupling) prefactors into a constant, we parametrize the effective low-frequency susceptibility amplitude as
\[
\chi_0(T) = \tilde{\chi}_0 \tanh \left(\frac{\beta E_{{\mathrm{res}}}}{2}\right),
\]
where $\tilde{\chi}_0$ is a phenomenological coefficient incorporating the coupling strengths, response time, TLS participation, and resonance lineshape. With our conventions, $\tilde{\chi}_0<0$ corresponds to positive damping and a negative frequency shift.

At large excitations, however, the response decreases with increasing drive as $1/\sqrt{ 1 + ( V/V_{{\mathrm{c}}} )^2 }$ which arises
due to the saturation of the resonant interaction. Consequently, we may write for the
resonant part of the relaxation rate as
\begin{equation}
    \label{eq:gamma_res}
        \Gamma_{{\mathrm{res}}} = \frac{\Gamma_{{\mathrm{res}},0}}{\sqrt{1 + \left( V/V_{{\mathrm{c}}} \right)^2}} \tanh \left( \frac{\beta E_{{\mathrm{res}}}}{2} \right).
\end{equation}
where $V_{\mathrm{c}}$ denotes the critical drive voltage and $\Gamma_{\mathrm{res},0}$ denotes the resonant-channel dissipation rate in the small-amplitude, zero-temperature limit. Alternatively, this suppression of the resonant interaction can be expressed in terms of 
the photon or phonon intensity, depending on the system coupled to the TLSs.

To obtain agreement with measured data, we presume that the
resonant coupling takes place between a subset of TLSs and a localized phonon mode whose boundaries are set by the electrode surface and the quartz fork, 
i.e. an out-of-plane $\lambda/4$ cavity-like mode of the electrode structure. The device geometry selects a characteristic
energy for this mode of around $E_{{\mathrm{res}}} \simeq 2 \pi \times \ 1.5\,$GHz \footnote{For Pd-based electrode with thickness 500 nm and sound velocity 3070 m/s, the frequency of transverse $\lambda/4$ mode is $E_{\mathrm{c}} \simeq 1.5$ GHz}. For a \qtyrange[range-phrase = --,range-units = single]{1}{1.5}{GHz} mode, the thermal
population is negligible at $T \lesssim 10\,$mK, becomes modest around \qtyrange[range-phrase = --,range-units = single]{20}{50}{mK}, and is clearly
non-negligible for $T \gtrsim$~\qtyrange[range-phrase = --,range-units = single]{50}{100}{mK}. However, the effective temperature of the fork's electrodes can exceed the bulk fork temperature 
due to absence of filtering between the cooled preamplifier gate terminal and the fork's electrodes. Owing to its mounting temperature ($\sim 4$\;K) and bias heating of 2.5 mW \cite{Zavjalov2019}, the HEMT channel operates at temperature $T \sim$~\qtyrange[range-phrase = --,range-units = single]{5}{10}{K} which can emit high-frequency noise through the gate capacitance.
Therefore, the phonon mode and some fraction of the TLSs may be populated even when the cryostat base temperature is in the few-mK range. Note that a large Kapitza boundary resistance between the electrodes and the fork allows for a large temperature difference between these two subsystems.

\paragraph{Temperature dependence of the relaxation channel.} In the case of the relaxational dissipation channel, we follow the standard tunneling model (see e.g.~\cite{Phillips.1987,Behunin2016}) and consider a mechanism in which TLSs couple
to strain by a deformation potential, essentially the strain modulating the TLS energy levels. 
For a low-frequency mechanical mode, the strain oscillates at a frequency much lower than the TLS separation. 
In the TLS eigenbasis this low-frequency drive is predominantly longitudinal and cannot resonantly 
flip the TLS. Instead, power is absorbed only because the population relaxes toward the instantaneously shifted equilibrium.

The temperature dependence of the relaxational channel is expected to be governed by a coupling to an ohmic bath, which yields relaxation rate 
$T_1^{-1} \propto E \coth \left[ E / 2 k_{\mathrm{B}} T \right] $. Assuming that the energies of the relevant TLSs are peaked around 
$E_{\mathrm{rel}}^*$, the dissipation rate arising from the relaxation interaction becomes 
\begin{equation}
    \label{eq:gamma_rel}
        \Gamma_{\mathrm{rel}} = \Gamma_{\mathrm{rel},0} \frac{\beta E_{\mathrm{rel}}^*}{\sinh (\beta E_{\mathrm{rel}}^*)} ,
\end{equation}
where $\Gamma_{\mathrm{rel},0}$ is a temperature-independent prefactor that sets the strength of the relaxation channel. A few details on the derivation of Eq.~\eqref{eq:gamma_rel} are given in App. \ref{sec:AppB}. Note that unlike the resonant interaction, this relaxation mechanism does not saturate at high intensities. App. \ref{sec:AppB} also compares the temperature dependencies given by Eqs.~\eqref{eq:gamma_res} and \eqref{eq:gamma_rel}.

The total dissipation rate is the sum of the electrical losses $\Gamma_{\mathrm{el}}$, linear mechanical dissipation $\Gamma_{\mathrm{m,bk}}$, and TLS contributions
\begin{equation}
    \label{eq:total_dissipation}
        \Gamma_{\mathrm{m,tot}} = 
        \Gamma_{\mathrm{el}} +
        \Gamma_{\mathrm{m,bk}} + \Gamma_{{\mathrm{res}}} + \Gamma_{\mathrm{rel}} .
\end{equation}
Ideally, $\Gamma_{\mathrm{el}}$ is linear, but in our setup there is a weak non-linearity in the gate capacitance $C_{\mathrm{g}}$ of the cooled preamplifier (see Appendix \ref{sec:characterization}). Because the non-linearity in $C_{\mathrm{g}}$ is negligibly small, we can lump $\Gamma_{\mathrm{el}}+\Gamma_{\mathrm{m,bk}}$ into a single linear dissipation $ \rightarrow \Gamma_{\mathrm{m,bk}}$. The detailed fitting process to the data is outlined in Appendix~\ref{sec:experiments}.

\section{Experimental setup}

\begin{center}
\begin{figure}[tb]
\centering
\includegraphics[width=0.6\linewidth]{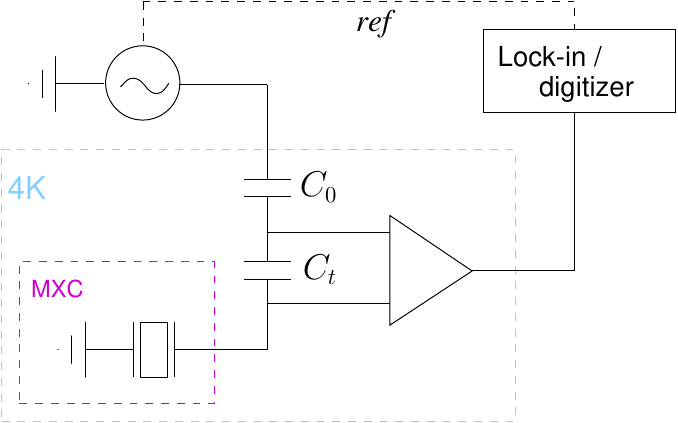}~~
\includegraphics[width=0.25\linewidth]{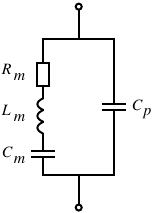}

\caption{a) Schematic illustration of the measurement circuit, with the temperature stages indicated by different colors. The readout capacitor $C_{\mathrm{t}}=300$\,pF converts the fork's terminal current into a voltage measured by the differential preamplifier; $C_0\gg C_{\mathrm{t}}$ acts as a DC block. b) Electrical equivalent circuit of the fork. The motional branch consists of $L_{\mathrm{m}}\simeq9550$\,H, $C_{\mathrm{m}}\simeq2.5$\,fF, and $R_{\mathrm{m}}\simeq1.6$\,k$\Omega$ at $T=4$\,mK. The parasitic capacitance $C_{\mathrm{p}}\simeq105$\,pF represents capacitive coupling across the fork terminals and shunts the motional branch. The bare motional frequency is $f_{\mathrm{bare}}=1/(2\pi\sqrt{L_{\mathrm{m}}C_{\mathrm{m}}})$, while the parallel resonance of the shunted fork is $f_{\mathrm{shunt}}=f_{\mathrm{bare}}\sqrt{1+C_{\mathrm{m}}/C_{\mathrm{p}}}$. Including $C_{\mathrm{t}}$ gives a detected resonance near $f_{\mathrm{detect}}\simeq f_{\mathrm{bare}}\sqrt{1+C_{\mathrm{m}}/(C_{\mathrm{p}}+C_{\mathrm{t}})}$, neglecting amplifier loading and the impedance of $C_0$. For our experimental parameters, the capacitive frequency shifts are small compared with the resonance frequency, so that $f_{\mathrm{detect}}\simeq f_{\mathrm{shunt}}\simeq f_{\mathrm{bare}}$. These fixed shifts are absorbed into the fitted resonance frequency; ideal capacitive loading does not change the amplitude-decay rate.}

\label{fig:Schematics}
\end{figure}
\end{center}

Our measurement circuit is shown in Fig.~\ref{fig:Schematics}a.
The mechanical resonator is a quartz tuning fork with dimensions of the two tines as: length $L=2.4$ mm, x-width $s \simeq 0.24$ mm, and y-width $w\simeq 0.15$ mm.
It has the regular tuning fork frequency $f=32768$ Hz ($=2^{15}$ Hz). The equivalent circuit of the fork is presented in Fig.~\ref{fig:Schematics}b.
One of the crystal electrodes is grounded to the cold plate of the cryostat allowing for better thermalization of the quartz crystal.
This allows the conduction electrons to cool effectively by electronic heat diffusion towards the electronic temperature of the copper refrigerant.

First, the quartz tuning fork is excited by a function generator either using continuous-wave (CW) output signal or using a short pulse of excitation at $f=32699$ Hz
which yields a broad enough frequency spectrum for the excitation of mechanical motion even with slightly shifted resonance frequency.
CW excitation was on for a time of $\sim 2Q/f_0$ (80 seconds or longer) to allow the resonator to reach nearly a steady-state response at each measurement frequency.

In the pulsed measurement scheme, the tuning fork oscillates freely at frequency $f_0$ and it generates an AC piezoelectric current, that is amplified by a home-made,
cooled voltage preamplifier in order to minimize interference from environmental noise and to maximize the conversion efficiency of the piezo-generated current.
The oscillating output voltage from the preamplifier is recorded using a Zurich Instruments UHFLI lock-in amplifier in capture mode using a sampling rate about 20 points per cycle (sampling rate $\sim 700$ kHz).
The decaying piezo signal is analyzed using a windowed Fourier transform over a sliding time spans of length $10^6$ points.
In continuous wave experiments, the output is detected by a lock-in amplifier SR830 using a time constant of 10 s.
Comparison of amplitudes in CW and pulsed excitation was made by direct comparison of the voltages across the fork in both cases.

In free decay experiments, the measurement bandwidth is on the order of 100 kHz, which results in a displacement sensitivity of $\langle \delta x \rangle_{\rm{rms}} \sim 0.1$ nm.
See Appendix~\ref{sec:characterization} for a more detailed overview of the device characterization and the measurement devices.

The resonance frequency and damping rate exhibit different temperature and amplitude dependences. These observations do not by themselves identify the spatial origin of the losses. However, two-level systems at surfaces or in disordered surface layers provide a plausible source of dissipation~\cite{Grajcar2017}, motivating the TLS-based interpretation developed here. COMMENT: the formula $\omega_{\mathrm{r}}=\omega_0\sqrt{1-1/Q^2}$ seems to have some issues. If it's ringdown, $1/Q^2$ should be $1/4Q^2$. I think that the sentence is in any case miselading, since the mechanism is irrelevant for our scales $\delta \omega \simeq 10^{-9} \mathrm{Hz}$.

\section{Experimental results}

We performed continuous-wave and free decay experiments on the quartz resonator in vacuum 
at temperatures $T =$~\qtyrange[range-phrase = --,range-units = single]{4}{1000}{mK}, where $T$ is the measured temperature of the refrigerator,
not accounting for possible overheating of the sample.
Our results displayed a strong drive amplitude dependence when the sample was cooled towards 
the base temperature, whereas very little influence of drive strength was observed at $T=1$ K.
In the linear excitation regime, we are able to determine the mechanical linewidth of the fork 
$\Delta f =\Delta \omega/2\pi \simeq 0.0260$ Hz (at $T \sim 1$\,K), corresponding a quality factor of $Q \simeq 12.6\cdot 10^5$.
The quality of our quartz resonators is similar to those employed for studies of quantum fluids.

\begin{center}
\begin{figure}[tb]
\centering
\includegraphics[width=0.95\linewidth]{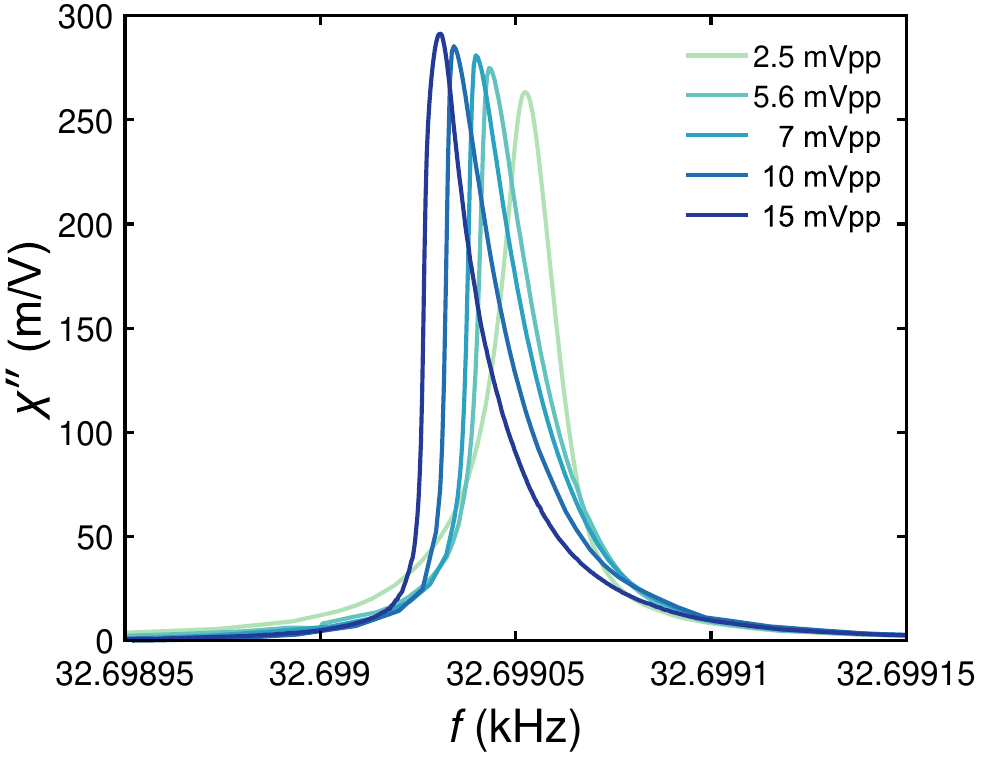}
\caption{Scaled absorption curves as a function of frequency recorded at temperature $T=4$~mK at drive voltages indicated in the inset (voltage after $C_0$ to ground).
The {``negative''} contribution in the non-linear dissipation is seen as an increase in the scaled amplitude of the Duffing shaped resonance signal. 
}
\label{fig:4mK}
\end{figure}
\end{center}

Driving the fork with increasing power reveals a ``softening'' spring mechanical non-linearity.
The response curve for drive-normalized absorption $\chi''(\omega)$ is displayed in Fig.~\ref{fig:4mK} 
as a function of drive power.
The data measured at $T=4$ mK indicate a clear increase in the maximum amplitude in $\chi''(\omega)$ at large drives.
This is in distinct contrast to typical behavior: constant scaled amplitude for pure Duffing 
behavior and decrease in the case of nonlinear dissipation~\cite{Lifshitz2008}.
Sometimes, e.g. in Ref.~\cite{Kumar2021}, there is an initial increase of the scaled 
mechanical response under continuous drive, which then quickly becomes dominated by increased nonlinear dissipation at larger drives.
In our case, the contribution of nonlinear dissipation remains ``negative'' and no decrease 
in scaled response is observed up to the highest drive voltage used (15 mV).
Furthermore, this behavior does not require continuous excitation, and it is seen also under free decay of oscillation.

The reduced dissipation was verified in free decay experiments of the resonance amplitude.
First, the fork was driven to steady-state oscillation using the same voltage amplitude as in the CW measurements.
The drive was then switched off and the decay of oscillation amplitude was followed as a function of time.
The time constant of the decay, $\tau_{\mathrm{d}}$, was obtained by an exponential fit to envelope curve of the decay.
Using the equation for decay rate $1/\tau = \pi \Delta f $, we obtained from low-amplitude decay rate $1/\tau \simeq 0.080 $ Hz a line width $\Delta f \simeq 0.0255 $ Hz, corresponding to $Q \simeq 12.8 \cdot 10^5$.
Thus, in the linear regime, the decay reproduces the $Q$ factor obtained from the CW measurements.
Similarly, the decrease in the line width at large drives is reproduced by the decay rate data $1/\tau$ of the envelope curve.
\begin{center}
\begin{figure}[tb]
\centering

\includegraphics[width=0.9\linewidth]{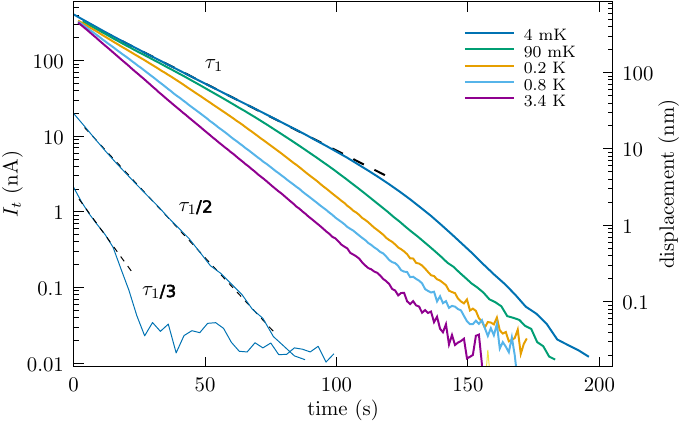}

\caption{Free decay of the oscillation amplitude of the fork measured at $T=4$\,mK after a strong excitation pulse which ramps the amplitude up to 570 nm (uppest trace, right scale).
The middle and bottom traces indicate the detection current amplitude (left scale) of the second and third harmonic signals, respectively, due to electrical nonlinearities. The overlaid lines indicate exponential decay with time constants $\tau_1$, $\tau_1/2$, and $\tau_1/3$ from top to bottom. }

\label{FD_4}
\end{figure}
\end{center}
The free ring-down of the fork's oscillation amplitude is illustrated in Fig.~\ref{FD_4}.
The initial decay is exponential, with time constant $\tau_1$, which starts to decrease around time $t = 100$ seconds.
In addition to the fundamental frequency $\omega_{\mathrm{\mathrm{FD}}}$ of the decay, we also display the decay of signals measured at $2\omega_{\mathrm{FD}}$ and $3\omega_{\mathrm{FD}}$.
Note that these are not modes of the fork but electrical signals that arise from nonlinear electrical behavior of the system during the decay at frequency $\omega_{\mathrm{FD}}$ (weakly non-linear amplifier gain with large input signals).
As expected, the second harmonic decay takes place with a time constant $\tau_1/2$.
It is also apparent from Fig.~\ref{FD_4} that the 3rd harmonic decays with time constant $\tau_1/3$.
The decay time constant $\tau_1$ agrees well with the $Q$ factor determined at equal amplitude using continuous drive.

\begin{center}
\begin{figure}[tb]
\centering

\includegraphics[width=0.95\linewidth]{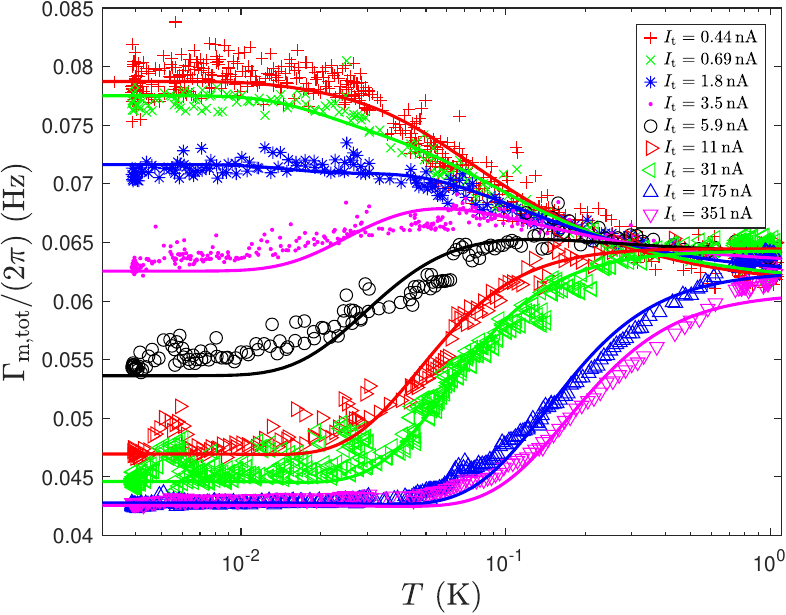}

\caption{Total relaxation rate measured from ring-down experiments at temperatures $T =$~\qtyrange[range-phrase = --,range-units = single]{4}{1000}{mK}. 
The inset denotes the oscillation amplitude in terms of the read-out current $I_{\mathrm{t}}$ for each of the nine data sets.
The overlaid curves are derived from Eq.~\eqref{eq:total_dissipation} using a procedure described in Appendix~\ref{sec:experiments}.
The resonant absorption is suppressed at high oscillation amplitude (large $I_{\mathrm{t}}$), leaving the relaxation absorption to dominate.}

\label{fig:Tdep}
\end{figure}
\end{center}

We followed the free decay of fork oscillations as a function of $T$ upto temperatures of 1\;K. In Fig.~\ref{fig:Tdep}, we compare Eq.~\eqref{eq:total_dissipation} to our measured results at nine different excitation amplitudes
given by the readout current (for conversion to oscillation amplitude, see Appendix~\ref{sec:characterization}).
The sequential fitting procedure, which first utilizes the importance of resonant relaxation at low $T$, is explained in Appendix~\ref{sec:experiments}.
In general, the fits agree very well with the data in Fig.~\ref{fig:Tdep}.
The fits suggest that the constant level splitting $E_{{\mathrm{res}}}^* = 2\pi \times 1.25\,$GHz of the TLSs taking part in the resonant coupling is justified. 
However, $E_{\mathrm{rel}}^*$ related to the relaxation mechanism depends quite strongly on the excitation voltage $V_{\mathrm{ex}}$ across capacitance $C_{\mathrm{t}}$ in Fig.~\ref{fig:Schematics},
(see Fig.~\ref{fig:7} in Appendix~\ref{sec:experiments}).
The approximately square-root voltage dependence is consistent with image-force-assisted lowering of the tunneling barriers of charged TLSs at the metal–quartz interface. Such barrier lowering enhances the tunneling matrix element and, at fixed asymmetry, increases the TLS splitting. Establishing this mechanism quantitatively would require a microscopic description of the interfacial potential and local electric field.

\begin{center}
\begin{figure}[b]
\centering

\includegraphics[width=0.95\linewidth]{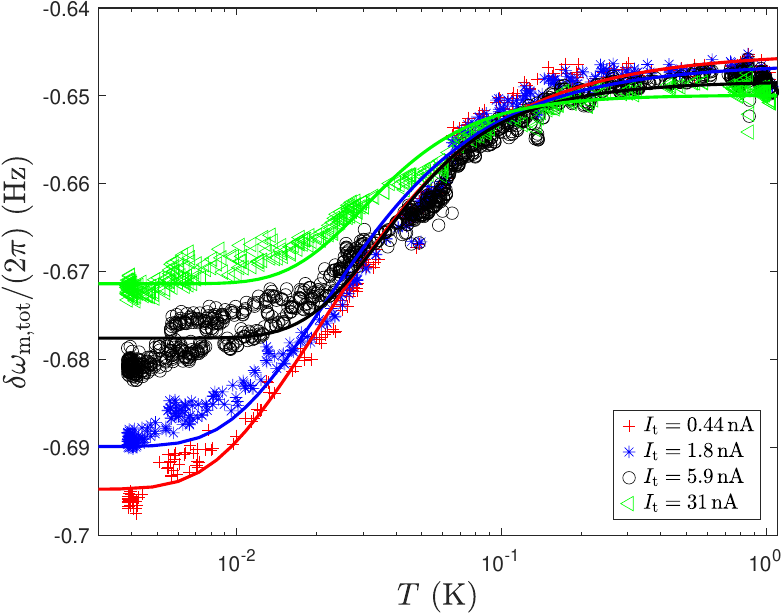}
\caption{The shift of the resonance frequency, 
$\delta \omega_{\mathrm{m,tot}} = 2\pi \times (f - f_0)$, 
in a ring-down measurement plotted as a function of $T$ at different, set read-out current values $I_{\mathrm{t}}$ given in the inset.
Here, the reference frequency is $f_0 = 32699.75\,$Hz.
The temperature dependence is first seen to decrease with increasing amplitude, which leads to crossing of curves.
}

\label{fig:freqTdep}
\end{figure}
\end{center}

Although free decay experiments have hundred-times worse displacement sensitivity than lock-in measurements, 
ring-down experiments are  better suited for studies of nonlinearities in the mechanical restoring force, which modify the resonant frequency of the fork.
The instantaneous frequency in the ring down versus oscillation amplitude can be used to determine the changes in the restoring force as a function of displacement.
Fig.~\ref{fig:freqTdep} displays the temperature dependence of the ring-down frequency $f$ at a constant oscillation amplitude.
The temperature dependence of $f$ is strongest 
at small amplitudes, and it is small at intermediate and very large amplitudes.

The fitted curves in Fig.~\ref{fig:freqTdep} are based on the TLS formalism discussed above for dissipation. The parametric resonant channel and the relaxation channel contributions can be applied to the change in frequency (cf. Eq.~\eqref{eq:6}). In the fits, we have employed forms of Eqs.~\eqref{eq:gamma_res} and \eqref{eq:gamma_rel} with the prefactors as fit parameters. The good agreement strengthens the applicability of our model to describe the non-linear dissipation phenomena with parametric backaction.

\section{Discussion}

The good agreement of our TLS model with experimental results on dissipation and frequency variations indicate that the non-linear phenomena arising from bulk properties of quartz or geometric non-linearities arising from geometry are not significant in our experiments. Since the quartz fork is of high quality, its bulk does not house many TLSs and most of the two level tunneling systems reside at the interface with the  disordered metallic electrodes. The validity of the TLS model indicates that the strain in the metallic electrodes, where the TLSs reside and couple to the mechanical modes in the electrode, can be considered small.

Usually at low temperatures, the quality of high-frequency resonances improves with increasing drive power.
In microwave cavities, large increase in the quality factor is observed with saturation of TLSs that are coupled (mostly resonantly) to the electric field in the cavity.
In our model, the low-frequency mechanical motion parametrically modulates the resonant coupling between TLSs and a microwave-frequency phononic mode of the metallic electrode. The relevant TLS splittings therefore lie in the microwave range. By contrast, TLSs resonant with the fundamental fork mode ($\sim 32\,$kHz) would have nearly equal ground- and excited-state populations at the experimental temperatures, strongly suppressing their net resonant absorption.

The metallic electrodes on the surface of the tuning fork may contain localized electromagnetic modes that are bounded by the electrode geometry. These modes, however, are at frequencies $f \sim c/NL$, where the speed of electromagnetic waves is $c\simeq 1.5 \times 10^8$\;m/s in quartz and $N=2$ (4) for a half-wave (quarter-wave) resonator. Using $L=2.4$\;mm, we estimate $f \sim 15$\;GHz and $f \sim 30$\;GHz for $\lambda/4$ and $\lambda/2$ resonators, respectively.
More relevant cavity frequencies are obtained for localized phononic modes. As the thickness of the metallic electrodes is about 500 nm, the out-of-plane, localized phonon modes are in the microwave regime ($f\approx 1-2 $\;GHz). 
Coupling these modes to TLSs, we obtain a channel for dissipation, for which the interactions with the TLSs are not saturated.
Since the coupling between strain and electric fields and two-level systems is well captured by the standard tunneling model~\cite{Black.1977}, the developed parametric back action model is conceptually well founded.

Ref.~\cite{Stockburger1995} investigated nonlinear acoustic response of low-energy excitation in the tunneling model for amorphous systems.
Their analysis reveals an increase \(Q \propto (\hbar\epsilon/k_{\mathrm{B}}T)Q_{\mathrm{lr}}\) where $\epsilon$ denotes 
the modulation of the TLS level splitting due to strain in the system and $Q_{\mathrm{lr}}$ is the quality factor in the linear regime.
The increase in $Q$ vanishes when $\hbar \epsilon \sim k_{\mathrm{B}} T$.
Using the TLS splittings obtained above in our model fitting,
the analysis of Stockburger et al.~\cite{Stockburger1995} parallels the findings in Fig.~\ref{fig:Tdep} 
for the low-excitation regime where $Q$ factor is found to increase from the linear regime value measured at high $T$.

Negative nonlinear dissipation (NDD) under continuous excitation has been carefully studied in recent years~\cite{Singh2016,Dykmann2019,Bousse2020,Ochs2022}.
In Ref.~\cite{Dykmann2019} it was coined as ``resonantly-induced frictional force'' (RIFF) in order to indicate that 
the phenomenon can drive the system even into self-sustained oscillations.
NDD (or RIFF) is typically assigned to an additional energy reservoir that can be employed for channeling power to the resonator motion.
For example, if there is drive-induced heating in the system, then thermal expansion may lead to additional driving component of the resonance 
and can cause an apparent decrease in the dissipation of the mechanical resonator.
It is noteworthy that in such a scenario, the drive has to be continuously connected in order to keep the dissipation at reduced level, 
and as such, would be consistent with our findings in the CW case.

In our system, however, the control mechanism of the NDD is more complex than in the resonantly-induced frictional force \cite{Dykmann2019}. In our quartz fork, microwave noise arising from a cooled preamplifier opens a new parametric dissipation mechanism which exhibits a strain-amplitude-dependent strength. When the amplitude of strain variation  is large (mechanical oscillation amplitude is large), the resonant part of this dissipation channel is saturated, in a similar fashion as happens with resonant TLSs in superconducting microwave cavities. The low-dissipation regime is sustained by the joint action of high-amplitude oscillation and the thermal noise of the preamplifier, which is distinct from the regular RIFF operation. Hence, our results indicate that extra high-frequency noise can lead in low frequency devices to effects which are clearly beyond a simple description by increase in temperature.


\section{Conclusions}

In conclusion, we investigate apparent negative nonlinear dissipation in quartz tuning forks down to few-mK temperature.
The dissipation rate in continuous-wave and ringdown measurements is highly dependent on the temperature and the driving of the fork, and the behavior is 
well described by combined contributions from standard linear mechanical dissipation and the tunneling two-level system model 
with resonant and non-resonant effects. The coupling to the TLSs is enabled by microwave phonon modes in the fork's electrodes.

\section*{Acknowledgments.} We are grateful to Mark Dykman, Jeevak Parpia, and Peter Skyba for discussions. This work was supported by the Academy of Finland (AF) projects 341913 (EFT), and 312295 \& 352926 (QTF). JM and FM acknowledge financial support from the Research Council of Norway (Grant No. 333937 and 36595) through participation in the QuantERA ERA-NET Cofund in Quantum Technologies. Our collaboration profited from the SuperQuMap COST action network. The research leading to these results has received funding from the European Union’s Horizon 2020 Research and Innovation Programme, under Grant Agreement No.~824109 (EMP). This work is part of the Finnish Centre of Excellence in Quantum Materials (QMAT). The experimental work benefited from the Aalto University OtaNano/LTL infrastructure.

\newpage

\appendix


\section{Experimental techniques and device characterization}
\label{sec:characterization}

In series with the fork, a ceramic capacitor suitable for cryogenic operation (C0G, Capacitance change 0 ppm per Grade C) is connected, 
serving as a read-out impedance of the current going through the fork (See Fig.~\ref{fig:Schematics}a).
The equivalent value of this capacitor was chosen to be $C_{\mathrm{t}}=300$ pF ($Z \simeq 16$ k$\Omega$ @32kHz).
Another capacitor $C_0=100$ nF  with a significantly smaller impedance was used as a dc-block.
The signal was measured across the readout capacitor using a cooled, high-impedance differential preamplifier~\cite{Zavjalov2019} with a slight modification:
the 50 $\Omega$ input resistance was replaced with a 10 M$\Omega$ resistor, ensuring that the input resistance does not load the readout circuit of the fork.
Two capacitors and the cooled preamplifier were placed in the liquid helium bath, right above the vacuum can top flange.
As the temperature of these components was kept constant, their temperature dependence could be excluded from contributing to non-linear behavior that we investigate in this work.

The equivalent circuit of a quartz fork is represented in Fig.~\ref{fig:Schematics}b.
It consists of a so-called motional branch shunted by a parasitic capacitance, $C_{\mathrm{p}}$.
The passive components $L_{\mathrm{m}}$, $C_{\mathrm{m}}$, and $R_{\mathrm{m}}$ are parameters corresponding to the basic 32.7 kHz resonance mode of our piezoelectric fork resonator 
(see e.g. Ref.~\cite{Blaauwgeers2007} and references therein).

To characterize the fork, we analysed its frequency response across the bare motional series resonance, $f_{\mathrm{bare}}$, and the parallel resonance of the capacitively shunted fork, $f_{\mathrm{shunt}}$, as defined below.
Using the equivalent model, we obtain for the mechanical inductance $L_{\mathrm{m}} = 9550$\,H 
and the resistance $R_{\mathrm{m}}=1.6$\,k$\Omega$ at resonance.
This yields for the piezoelectric coefficient $a=3.4$\,$\mu \mathrm{C/m}$ for the given dimensions of the fork, 
which is close to the values obtained on similar forks (see e.g. Ref.~\cite{Schomeranzer2011}).
The equivalent model can also be used for deriving the linear-regime relation between oscillation amplitude 
and the voltage at the input of the preamplifier (i.e. across capacitance $C_{\mathrm{t}}$).
The piezoelectric coefficient can be written as $a=I_{\mathrm{t}}/v$ where $v$ is the effective velocity of the tines 
and $I_{\mathrm{t}}$ is the current through the read-out capacitor: $I_{\mathrm{t}} = V_{\mathrm{t}}/(1/\omega C_{\mathrm{t}})$ with $V_{\mathrm{t}}$ across the read-out capacitor.
Having preamplifier noise $\delta V_t$, this yields for the sensitivity of motion $\delta x=\delta V_{\mathrm{t}} C_{\mathrm{t}}/a$.
This relation corresponds to a resolution of 0.3 pm in a lock-in measurement using a time constant $\tau = 1\,$s 
and a preamplifier with input noise of 3 nV/Hz$^{1/2}$~\cite{Zavjalov2019}.

The detected resonance frequency, $f_{\mathrm{detect}}$, can also shift slightly with drive amplitude due to non-linear gate capacitance of the HEMT-based preamplifier. Ignoring $R_{\mathrm{m}}$, the motional branch impedance is
$Z_{\mathrm{m}}=j\omega L_{\mathrm{m}}+\frac{1}{j\omega C_{\mathrm{m}}}$.
Its admittance is $Y_{\mathrm{m}}=\frac{1}{Z_{\mathrm{m}}}$, while
for the shunt capacitance $C_{\mathrm{p}}$, we have
$Y_{\mathrm{p}}=j\omega C_{\mathrm{p}}$.
At the parallel resonance of the shunted fork, $\omega_{\mathrm{shunt}}=2\pi f_{\mathrm{shunt}}$, the total susceptance is zero:
$\operatorname{Im}(Y_{\mathrm{m}})+\operatorname{Im}(Y_{\mathrm{p}})=0$.
This yields
\begin{equation}
\omega_{\mathrm{shunt}}=
\frac{1}{\sqrt{L_{\mathrm{m}}}} \sqrt{\frac{1}{C_{\mathrm{m}}}+\frac{1}{C_{\mathrm{p}}}}.
\end{equation}
The shunted-fork frequency relative to the bare motional frequency, $f_{\mathrm{bare}}=1/(2\pi\sqrt{L_{\mathrm{m}} C_{\mathrm{m}}})$, is given by
\begin{equation}
\frac{f_{\mathrm{shunt}}}{f_{\mathrm{bare}}}=\sqrt{1+\frac{C_{\mathrm{m}}}{C_{\mathrm{p}}}}.
\end{equation}
Neglecting amplifier loading and the impedance of $C_0$, the resonance detected across $C_{\mathrm{t}}$ occurs near
\begin{equation}
f_{\mathrm{detect}}\simeq f_{\mathrm{bare}}\sqrt{1+\frac{C_{\mathrm{m}}}{C_{\mathrm{p}} + C_{\mathrm{t}}}}.
\end{equation}
The fixed capacitive frequency shifts are small compared with the resonance frequency and are absorbed into the fitted center frequency; ideal capacitive loading leaves the amplitude-decay rate unchanged.
Using $C_m=2.5$\;fF and $C_p=105$\;pF, we obtain $(f_{\mathrm{shunt}}-f_{\mathrm{bare}})/f_{\mathrm{bare}}\simeq12$\;ppm, corresponding to $f_{\mathrm{shunt}}-f_{\mathrm{bare}}\simeq0.4$\;Hz. With $C_{\mathrm{t}}=300$\;pF, the detected resonance is shifted by $f_{\mathrm{detect}}-f_{\mathrm{bare}}\simeq0.10$\;Hz. Taking an estimate of capacitive nonlinearity from Ref. \footnote{\href{https://www.microwavejournal.com/articles/11108-a-hemt-large-signal-model-with-improved-transconductance-and-gate-capacitance-peaking-characteristics}{A HEMT large signal model with improved transconductance and gate capacitance peaking characteristic}, Microwave Journal (2011).}, $dC_g/dV_g=0.3$\;pF/V, we obtain a capacitance change $\delta C_p \simeq 3$\;fF for 10\;mV$_{p-p}$ variation on the gate. This corresponds to at most $\delta f_{\rm{non-linear}} \sim 10$\;$\mu$Hz for a spurious frequency change due to non-linearity in the preamplifier's gate capacitance. This is less than the scatter in the measured data points.

The experiments were performed on a refrigerator based on a Leiden Cryogenics DR2000 dilution refrigerator~\cite{Yao2000}, with a minimum mixing chamber temperature of 2.8 mK.
The temperature of the nuclear stage was measured using a resistor calibrated against a noise thermometer; Pt NMR thermometry was also used at the lowest temperatures.
Despite the good thermal anchoring of the fork to the nuclear stage, the sample may overheat by a strong drive at the lowest temperatures.
We estimate the maximum heating at resonance to be $<2\,$pW using 1.6 k$\Omega$ for the equivalent resistance of the fork.
Using typical values for electron-phonon coupling $\Sigma \simeq 10^9$ W/K$^5$m$^3$ ($\dot{Q}_{e-ph} =\Sigma(T_e^5-T_0^5)$),
this heating power will limit temperatures to $T_e \simeq 40$ mK at the largest drives 
(assuming the power is uniformly distributed to the electrical leads with volume $2\times 10^{-14}$ m$^3$ with no substantial overheating of the phonons).

In the linear excitation regime, the shape of the mechanical response curve 
can be fitted by absorption and dispersion formulas given by
\begin{eqnarray}
    \label{eq:resp}
        \chi ''(\omega ) = \frac{{{I_0}{{(\Delta \omega )}^2}{\omega ^2}}}{{{{(\Delta \omega )}^2}{\omega ^2} + {{({\omega ^2} - \omega_{\mathrm{detect}}(T)^2)}^2}}},\\
        \nonumber \chi '(\omega ) = \frac{{{I_0}\Delta \omega \omega ({\omega ^2} - \omega_{\mathrm{detect}}(T)^2)}}{{{{(\Delta \omega )}^2}{\omega ^2} + {{({\omega ^2} - \omega_{\mathrm{detect}}(T)^2)}^2}}},
\end{eqnarray}
respectively.
Here, the prefactor $I_0$ depends on the size of the tuning fork and the properties 
of quartz~\cite{Blaauwgeers2007}, $\omega_{\mathrm{detect}}(T)=2\pi f_{\mathrm{detect}}(T)$ denotes the temperature-dependent center frequency of the detected resonance, and $\Delta \omega$ stands for the full width of 
the absorption curve.
The agreement of Eqs.~\eqref{eq:resp} to our data are excellent in linear regime, 
and their fits can be used to accurately determine the quality factor.
We find for the resonance linewidth $\Delta f =\Delta \omega/2\pi \simeq 0.0260$ Hz (at $T \sim 1$\,K), 
which corresponds to a quality factor of $Q \simeq 12.6 \cdot 10^5$ as mentioned in the main text.

\begin{center}
\begin{figure*}[h!tb]
\centering

\includegraphics[width=0.95\linewidth]{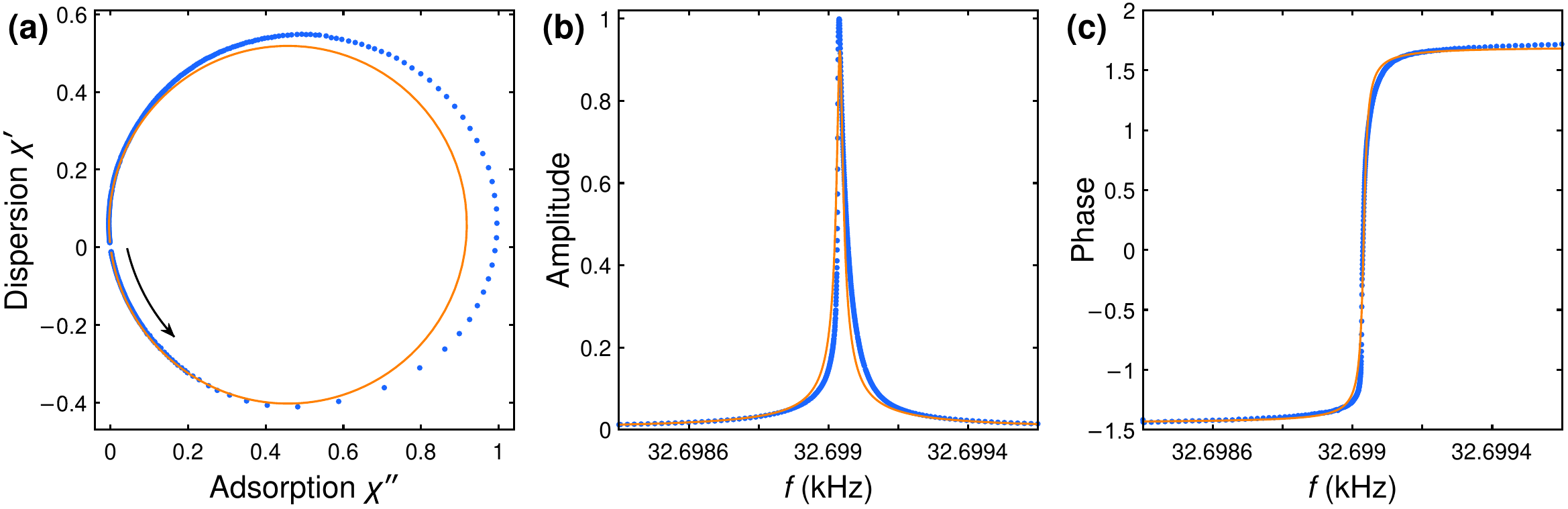}

\caption{Absorption (its maximum scaled to one) and dispersion responses of the fork measured at a large drive voltage of 15 mV at capacitance $C_{\mathrm{t}}$, see Fig.~\ref{fig:Schematics}.
a) Parametric plot of dispersion $\chi'$ against $\chi''$ with increasing frequency running in counterclockwise fashion along the nearly circular trace starting from (0,0).
b) Resonance magnitude as a function of drive frequency, showing a slight deviation of the fit from data points towards lower frequencies (spring softening).
c) Phase of the data with a fit curve obtained using Eqs.~\eqref{eq:resp}.
The same equations were used in fitting the magnitude data. }

\label{fig:duffingData}
\end{figure*}
\end{center}

With increasing drive, the measured resonance shape starts to deviate from the linear regime, 
which is illustrated in Fig.~\ref{fig:duffingData}.
Figure~\ref{fig:duffingData}a shows that the response circle (parametric plot of $\chi '(\omega )$ vs. $\chi ''(\omega ))$ 
arising from Eqs.~\eqref{eq:resp} does not catch the full extent of the data.
Measuring the half-height width, we obtain smaller width than at smaller drive.
This indicates reduced dissipation, which could be an indication of a hidden, extrinsic drive channel, 
which appears to compensate part of the dissipation using energy from the drive.
In Figs.~\ref{fig:duffingData}a and~\ref{fig:duffingData}b, we display fits of Eqs.~\eqref{eq:resp} to magnitude 
$\chi(\omega )=\sqrt{(\chi ''(\omega ))^2+\chi '(\omega)^2}$ and phase $\theta = \arctan (\chi '(\omega )/\chi ''(\omega ))$, respectively.
The fit of the resonance magnitude displays a small deviation from data points towards lower frequencies.
This indicates that the mechanical non-linearity is of ``softening spring'' type.

\section{Modeling dissipation}
\label{sec:AppB}

\subsection{Relaxation absorption}
\label{sec:models3}

On top of the resonant absorption described above in the main text, there is an additional off-resonant contribution,
relaxation absorption, that arises as the TLS energy levels are modified due to the TLSs interacting with phonons.
For a mechanical mode, the strain oscillates at a frequency much lower than the TLS separation.

We follow the standard TLS model having a bath of tunneling two-level states with a distribution $P(\Delta, \Delta_0)$, each of them described by 
$H_{{\mathrm{TLS}}} = \frac{1}{2}\left(\Delta \sigma_z + \Delta_0 \sigma_x\right)$. We employ the shorthand notation 
$g_{\parallel} = \Delta / E$ and $g_{\perp} = \Delta_0 / E$ with $E = \sqrt{\Delta^2 + \Delta_0^2}$.

For averaging over the TLS distribution in Fermi–Golden–Rule analysis, we assume a TLS distribution that is uniform in $\Delta$ and log-uniform in $\Delta_0$, 
$P(\Delta , \Delta_0) = P_0 / \Delta_0$. We obtain 
\begin{equation}
    \label{eq:gamma_rel_derivation_1}
        \Gamma_{\mathrm{rel}} = C_{\mathrm{rel}} \int d \Delta \, d \Delta_0 \, P(\Delta,\Delta_0) \, g_{\parallel} (\Delta,\Delta_0)^2 \, 
        \frac{\partial m_{\mathrm{eq}}}{\partial E} \, F(\Omega_{\mathrm{m}},T_1),
\end{equation}
where
\begin{subequations}
    \begin{align}
        \frac{\partial m_{\mathrm{eq}}}{\partial E} &= \frac{\partial}{\partial E} \tanh \left(\frac{\beta E}{2}\right)
            = \frac{\beta}{2} \, \cosh^{-2} \left(\frac{\beta E}{2}\right) , \\
        F(\Omega_{\mathrm{m}},T_1) &= \frac{\Omega_{\mathrm{m}} T_1(E,\Delta_0,T)}{1 + \Omega_{\mathrm{m}}^2 T_1 (E,\Delta_0,T)^2}, 
    \end{align}
\end{subequations}
are the ``energy susceptibility'' of the TLSs equilibrium population and the
Debye factor, respectively, characterizing the out-of-phase response of the system to a
perturbation.

We assume that the relaxation time of the TLSs taking part in the relaxation interaction to be 
consistent with a description in terms of relaxation to an ohmic bath.
\begin{equation}
    \label{eq:ohmic_bath_T1}
        T_1^{-1}(E,T)= A_{\mathrm{ph}}\, g_{\perp}^2\, E\, \coth \left( \frac{\beta E}{2} \right),
\end{equation}
where $A_{\mathrm{ph}}$ collects material and geometry constants. 
In the "frozen" limit $\Omega_{\mathrm{m}} T_1 \gg 1$ and for a narrow $E$‑band,
\begin{equation}
    \label{eq:8}
        F(\Omega_{\mathrm{m}},T_1)  \simeq \frac{1}{\Omega_{\mathrm{m}} T_1} \;\simeq\; \frac{A_{\mathrm{ph}}}{\Omega_{\mathrm{m}}} \, 
            g_{\perp}^2\, E\, \coth \left( \frac{\beta E}{2} \right) .
\end{equation}
For the distribution $P(\Delta,\Delta_0)=P_0/\Delta_0$, introduce $(E,u)$ through $\Delta=Eu$ and $\Delta_0=E\sqrt{1-u^2}$. Taking $\Delta\in(-\infty,\infty)$ and $\Delta_0>0$ gives $E>0$ and $-1<u<1$. The absolute Jacobian is $E/\sqrt{1-u^2}$, so that
\[
d\Delta\,d\Delta_0=\frac{E}{\sqrt{1-u^2}}\,dE\,du.
\]
With $P=P_0/[E\sqrt{1-u^2}]$, $g_{\parallel}^2=u^2$, and $g_{\perp}^2=1-u^2$, the product of the distribution, coupling factors, and Jacobian reduces to $P_0u^2$. The $u$ integration therefore gives $\int_{-1}^{1}u^2\,du=2/3$. This constant is absorbed into the overall prefactor, yielding
\begin{equation}
    \label{eq:gamma_rel_derivation_2}
        \Gamma_{\mathrm{rel}}
            \propto \int dE \frac{E}{k_{\mathrm{B}} T}\; \frac{1}{\sinh \left(\frac{E}{ k_{\mathrm{B}} T}\right)}.
\end{equation}
We assume that the TLSs contributing appreciably to relaxation absorption have a narrow distribution of splittings around $E_{\mathrm{rel}}^*$. We therefore evaluate the thermal factor at $E = E_{\mathrm{rel}}^*$ and absorb the integrated weight of the participating TLS distribution into $\Gamma_{\mathrm{rel},0}$, yielding
\begin{equation}
    \label{eq:gamma_rel_appendix}
        \Gamma_{\mathrm{rel}} = \Gamma_{\mathrm{rel},0} \frac{ \beta E_{\mathrm{rel}}^*}{\sinh(\beta E_{\mathrm{rel}}^*)} , 
\end{equation}
where all constants have been absorbed in the prefactor $\Gamma_{\mathrm{rel},0}$.

Unlike the dissipation arising from the resonant interaction, the relaxation interaction does not saturate at the high-intensity limit. 
On the other hand, the relaxation contribution to the total dissipation is insignificant at low temperatures.

\subsection{Comparison of resonant and relaxational dissipation}
\label{sec:models4}

The total dissipation influencing the quartz forks reads
\begin{equation}
    \label{eq:total_dissipation_appendix}
    \begin{split}
        \Gamma_{\mathrm{m,tot}} =& \Gamma_{\mathrm{m,bk}} 
            + \frac{\Gamma_{{\mathrm{res}},0}}{\sqrt{1 + \left( V/V_{{\mathrm{c}}} \right)^2}} \tanh \left( \frac{\beta E_{{\mathrm{res}}}^*}{2} \right) \\
            &+ \Gamma_{\mathrm{rel},0} \frac{ \beta E_{\mathrm{rel}}^*}{\sinh(\beta E_{\mathrm{rel}}^*)}
    \end{split}
\end{equation}
with distinct contributions from the three effects detailed above.
The amplitude and temperature dependence is rather involved, 
and it yields apparent negative differential dissipation for the mechanical motion.
However, it is worth highlighting that all the dissipative contributions are positive.

\begin{figure}[htb]
    \centering
    \includegraphics[width=0.9\linewidth]{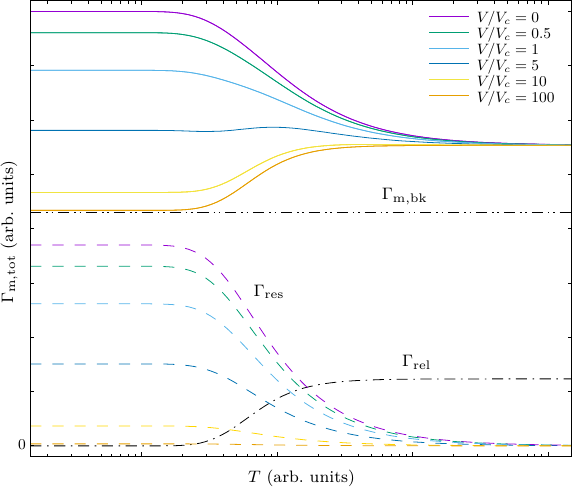}
    \caption{Solid curves display the temperature dependence of the total dissipation 
    $\Gamma_{\mathrm{m,tot}}=\Gamma_{\mathrm{m,bk}}+\Gamma_{{\mathrm{res}}}+\Gamma_{\mathrm{rel}}$ at six excitation amplitudes ($I^{\mathrm{ph}}/I_{{\mathrm{c}}}^{\mathrm{ph}} \propto (V/V_{{\mathrm{c}}})^2$). 
    The contributions of linear background, resonant, and relaxation damping to the total dissipation are shown by the dashed traces.
    The apparent negative dissipation can be modeled using Eq.~\eqref{eq:total_dissipation_appendix}, 
    where different terms dominate or are suppressed at different temperatures and oscillation amplitudes.}
    \label{fig:2}
\end{figure}

The temperature dependence of the total dissipation and its individual 
components is qualitatively illustrated in Fig.~\ref{fig:2}
at a few excitation amplitudes $A$ driving the system beyond the critical 
threshold specified by a critical voltage ($V_{\mathrm{ex}}/V_{{\mathrm{c}}} = 0, 0.5, 1, 5, 10, 100$)
corresponding to a certain phonon flux $I_{{\mathrm{c}}}^{\mathrm{ph}}$ . 
The apparent negative dissipation contribution, seen here in the $\Gamma_{\mathrm{m,tot}}(T)$ data, 
is due to the different effects being significant or suppressed in different regimes in the $(T,A)$ plane.
In particular, relaxation absorption is insignificant at low temperatures, contrary to resonant absorption 
that saturates at high $T$. Focusing on the low-$T$ regime, we can thus isolate the resonant absorption 
contribution versus the drive voltage from the measurements (2nd term in Eq.~\eqref{eq:total_dissipation_appendix}).
Similarly, when the fork is driven to a large amplitude $A$, the resonant absorption mechanism becomes suppressed, 
and we can see the relaxation effect on its own, with increased dissipation as the temperature increases (3rd term in Eq.~\eqref{eq:total_dissipation_appendix}).
We can thus conclude that, taking the low temperature and high excitation limits of the data, 
we can reveal the underlying background linear dissipation rate, as both TLS mechanisms are suppressed in this case.

\section{Extraction of parameters and the fitting process}
\label{sec:experiments}

\subsection{Dissipation contributions}
\label{sec:dissipation_fits}

Let us first focus on the low temperature regime to isolate the resonant TLS absorption.
Looking at Eq.~\eqref{eq:total_dissipation_appendix}, we see that at very low temperatures 
(we choose $4\,$mK here, and assume $\frac{E_{\mathrm{rel}}^*}{2 k_B T} \gg 1$),
this simplifies to 
$\Gamma_{\mathrm{m,tot}} \approx \Gamma_{\mathrm{m,bk}} + \Gamma_{{\mathrm{res}},0} /\sqrt{1 + (V_{\mathrm{ex}}/V_{{\mathrm{c}}})^2}$, 
which we fit to the data with $\Gamma_{\mathrm{m,bk}}$, $\Gamma_{{\mathrm{res}},0}$, and $V_{{\mathrm{c}}}$ as the fitting parameters.

\begin{figure}[htb]
    \centering
    \includegraphics[width=0.9\linewidth]{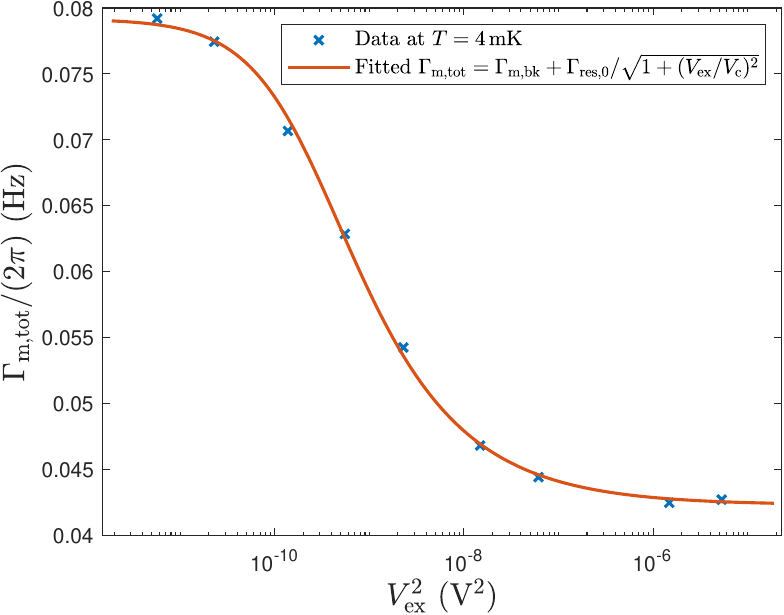}
    \caption{Total dissipation rate at $T = 4\,$mK as a function of the square of the excitation voltage $V_{\mathrm{ex}}$ across capacitance $C_{\mathrm{t}}$. 
    }
    \label{fig:4}
\end{figure}

Looking at the fit results in Fig.~\ref{fig:4}, we see that we obtain the same behavior 
as discussed in the TLS literature on glasses at low $T$~\cite{Golding.1976}, 
sometimes expressed in terms of the intensity of the phonon field.

We obtain the linear dissipation rate 
$\Gamma_{\mathrm{m,bk}} = 2\pi \times 0.042\,$Hz,  
the prefactor for the resonant dissipation process 
$\Gamma_{{\mathrm{res}},0} = 2\pi \times0.037\,$Hz, and
the critical voltage
$V_{{\mathrm{c}}} \approx 1.54\times10^{-5}\,$V. 
At this critical excitation, resonant dissipation has been suppressed by $1/\sqrt{2}$, which occurs in the center of our excitation voltage range.
Hence, we may clearly access both the unsuppressed and fully suppressed regimes as needed.

Using the dissipation parameters obtained above, we can now apply the full
model, Eq.~\eqref{eq:total_dissipation_appendix}, to fit the temperature and amplitude dependence of the data.
We fix the energy separation of the TLS taking part in the resonant process at 
$E_{\mathrm{res}}^* = 2\pi \times 1.25 \,$GHz that is motivated by the device geometry as it is 
consistent with a standing microwave mode whose boundaries are set by the electrode surfaces, 
i.e. an out-of-plane cavity-like mode.
The remaining fitting parameters needed are $\Gamma_{\mathrm{rel},0}$ and $E_{\mathrm{rel}}^*$ 
describing the relaxation mechanism that was absent in the low-temperature data.
The results of fitting the full model, Eq.~\eqref{eq:total_dissipation_appendix}, to the data are shown in Fig.~\ref{fig:Tdep} in the main text.
We obtain $\Gamma_{\mathrm{rel},0} \approx 2\pi \times 0.02\,$Hz, however, $E_{{\mathrm{rel}}}^*$ exhibits a strong $\sqrt{V}$-dependence.

\begin{figure}[htb]
    \centering
    \includegraphics[width=0.9\linewidth]{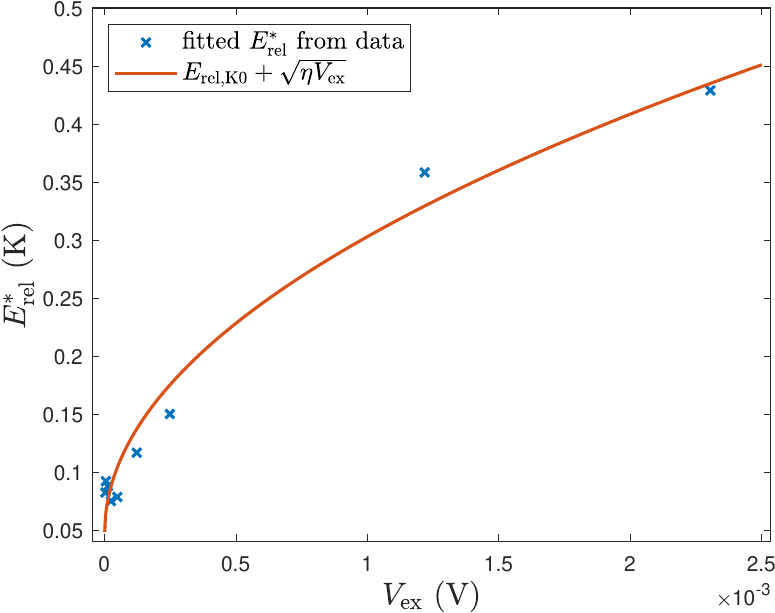}
    \caption{The voltage dependence of the energy of the TLSs taking part in relaxation absorption, $E_{\mathrm{rel}}^*$.}
    \label{fig:7}
\end{figure}

Let us finally take a look at this voltage dependence of $E_{\mathrm{rel}}^*$. 
In Fig.~\ref{fig:7}, we see that fitting 
$E_{\mathrm{rel}}^* = E_{\mathrm{rel},\mathrm{K}0} + \sqrt{\eta V_{\mathrm{ex}}}$
to the energy values extracted from the full temperature dependence fits (Fig.~\ref{fig:Tdep}) 
gives a reasonably good agreement with 
$E_{\mathrm{rel},\mathrm{K}0} \simeq 0.048\, \mathrm{K} = 2\pi \times 1.0\,$GHz, a value close to $E_{{\mathrm{res}}}^*$, and 
$\eta = 64.9\,$K$^2$/V.

\subsection{Frequency shift contributions}
\label{sec:frequency_shift_fits}

Fitting to the frequency shift data follows the same sequential process as for 
the dissipation phenomena, with the advantage that the critical voltage, 
$V_{\mathrm{c}}$, determining the suppression of the resonant component of 
the frequency shift $\delta \omega_{\mathrm{res}}$, is readily obtained 
from the dissipation analysis.

\begin{figure}[htb]
    \centering
    \includegraphics[width=0.9\linewidth]{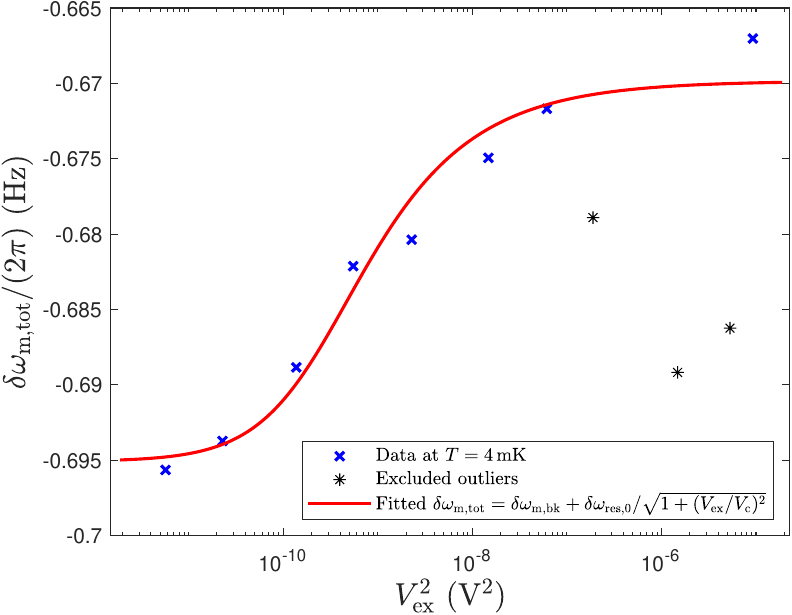}
    \caption{Total frequency shift $\delta \omega_{\mathrm{m,tot}} = 2\pi \times (f - f_0)$ at $T = 4\,$mK as a function of the square of the excitation voltage $V_{\mathrm{ex}}$. The reference frequency is $f_0 = 32699.75\,$Hz.
    }
    \label{fig:8}
\end{figure}

The frequency shift obeys the expected behavior from our TLS model, as seen 
from the low temperature analysis in Fig.~\ref{fig:8}. The results obtained 
from this fit are carried over to the fitting of the full model as a function 
of temperature, as shown in Fig.~\ref{fig:freqTdep} of the main text. Note that 
there some measurements at certain excitation voltages are omitted for visual
clarity.

This brief analysis of the frequency shift, arising from the interactions 
with TLSs, is in agreement with the data and further supports our explanation
of the nonlinear dissipation of the quartz fork using the tunneling TLS model.

\bibliography{bibliography.bib}

\end{document}